\documentclass[aps,prb,preprint,preprintnumbers,amsmath,amssymb,floatfix,superscriptaddress]{revtex4-1}
\usepackage[utf8]{inputenc}
\usepackage[T1]{fontenc}
\usepackage[
  top=8mm,
  bottom=7mm,
  left=1.5cm,
  right=1.5cm,
  headheight=14pt,
  includehead,includefoot,
  heightrounded, 
]{geometry} 
\usepackage{color}
\usepackage{amsmath}
\usepackage{amsfonts}
\usepackage{hyperref}
\usepackage{comment}
\usepackage{multirow}
\usepackage{amsthm}
\usepackage{bm}
\usepackage{float}
\usepackage{fancyhdr}
\usepackage{filecontents}
\hypersetup{
  colorlinks = true,
  allcolors= blue,
}
\usepackage[percent]{overpic}
\usepackage{tikz,pgfplots}
\usepackage{blkarray}
\usepackage[normalem]{ulem}
\usepackage{bibunits}
\usepackage{etoolbox}
\usepackage{mathrsfs}

\makeatother
\DeclareUnicodeCharacter{300}{à}

\DeclareMathAlphabet{\mathsfit}{\encodingdefault}{\sfdefault}{m}{sl}
\SetMathAlphabet{\mathsfit}{bold}{\encodingdefault}{\sfdefault}{bx}{sl}
\newcommand{\tens}[1]{\bm{\mathsfit{#1}}}
\newcommand{\tenscomp}[1]{\mathsfit{#1}}
\makeatletter
\fancypagestyle{footmark}{
  \fancyfoot[L]{\footmark}
}

\makeatother

\makeatletter
\renewcommand*{\@fnsymbol}[1]{\ifcase#1\else\@arabic{\numexpr#1\relax}\fi}
\makeatother
\makeatletter
\newcommand*{\newbibstartnumber}[1]{%
  \apptocmd{\thebibliography}{%
    \global\c@NAT@ctr #1\relax
    \addtocounter{NAT@ctr}{-1}%
  }{}{}%
}
\makeatother
\begin{document}
\title{ \textbf{A note on a unified theory of thermal transport in crystals and disordered solids}}
\author{ {{Samuel Huberman}}}
\affiliation{Department of Chemical Engineering, McGill University, 
Montreal, Quebec H3A 0C5, Canada}
\begin{abstract}
We propose an extension to the original result derived by Simoncelli et al.~\cite{simoncelli2019unified} to encompass the effects of a space and time dependent heat source. Via a Fourier Transfrom, we obtain closed form expressions for the dynamics of the phonon population and coherence as a function of heat source.
\end{abstract}
\maketitle
We propose an extension to the original result derived by Simoncelli et al.~\cite{simoncelli2019unified} to encompass the effects of a space and time dependent heat source. This approach has been successfully applied to study the size effects in silicon germanium alloys~\cite{huberman2017unifying} and phonon hydrodynamics~\cite{huberman2019observation}, but there is a missing piece for equivalent work in non-crystalline solids. We seek to address this gap. For ease, we use the same notation as ~\cite{simoncelli2019unified}.

\begin{widetext}
\begin{equation}
\frac{\partial  }{\partial  t} \tens{N}({\bm{R}},{\bm{q}},t)
+i\Big[\bm{\Omega}(\bm{q}),\tens{N}(\bm{R},\bm{q},t)\Big]+\frac{1}{2}\Big\{\vec{\tens{V}}(\bm{q}),\cdot \vec{\nabla}_{\bm{R}} \tens{N}(\bm{R},\bm{q},t)\Big\}=\frac{\partial }{\partial  t}\tens{N}(\bm{R},{\bm{q}},t)   \bigg|_{\mathrm{H}^{\rm col} } + \tens{P}({\bm{R}},{\bm{q}},t), \label{eq:Wigner_evolution_equation_N}
\end{equation}
\end{widetext}

where $\tens{N}({\bm{R}},{\bm{q}},t)$ is the generalization of a phonon distribution (from a rank 1 to a rank 2 tensor), $\vec{\tens{V}}(\bm{q})$ is a generalization of a phonon's group velocity and $\Omega(\bm{q})_{s,s'}{=}\omega(\bm{q})_s\delta_{s,s'}$. We have added the term $\tens{P}({\bm{R}},{\bm{q}},t)$ which describes the heat source. For the purposes of generality, $\tens{P}({\bm{R}},{\bm{q}},t)$ is a matrix-valued function, where the diagonal terms correspond to the production of phonon population and the off-diagonal terms allow for the production of phonon coherence\footnote{A carefully designed experiment that is capable of generating populations of distinct phonon modes in a correlated way might be one approach.}. To solve Eqn.~\ref{eq:Wigner_evolution_equation_N}, we take the following steps. Fourier transforming Eqn.~\ref{eq:Wigner_evolution_equation_N} where $\omega$ and $\bm{K}$ are the temporal and spatial Fourier variables:

\begin{widetext}
\begin{equation}
-i \omega \tens{\tilde{N}}({\bm{K}},{\bm{q}},\omega)
+i\Big[\bm{\Omega}(\bm{q}),\tens{\tilde{N}}({\bm{K}},{\bm{q}},\omega)\Big]+\frac{1}{2}\Big\{\vec{\tens{V}}(\bm{q}),\cdot -i\bm{K} \tens{\tilde{N}}({\bm{K}},{\bm{q}},\omega)\Big\}=\mathscr{F} \Big\{\frac{\partial }{\partial  t}\tens{N}({\bm{R}},{\bm{q}},t)  \bigg|_{\mathrm{H}^{\rm col} }\Big\} + \tens{P}({\bm{K}},{\bm{q}},\omega), \label{eq:Wigner_evolution_equation_N_FT}
\end{equation}
\end{widetext}

where linearizing about about the equilibrium temperature allows for a closed form expression for the collision term:

\begin{equation}
\medmuskip=0mu
\thinmuskip=0mu
\thickmuskip=0mu
\begin{split}
\mathscr{F} &\Big\{\frac{\partial }{\partial  t}\tens{N}({\bm{R}},{\bm{q}},t)  \bigg|_{\mathrm{H}^{\rm col} }\Big\}{=}\hspace*{1mm}-(1-\delta_{s,s'})\frac{\Gamma(\bm{q})_{s}+\Gamma(\bm{q})_{s'}}{2} \tenscomp{\tilde{N}}(\bm{K},\bm{q},\omega)_{s,s'}\\
 &{-}\frac{\delta_{s,s'}}{\mathcal{V}N_{\rm c}}\hspace*{-0.5mm}{\sum\limits_{{s''}{\bm{q}''}}}\hspace*{-0.5mm} \tenscomp{{A}}^T(\bm{q},{\bm{q}''})_{\hspace*{-0.4mm}s,{s''}}\hspace*{-0.5mm}\big(\hspace*{-0.5mm}\tenscomp{\tilde{N}}(\bm{K},{\bm{q}''}\hspace*{-1mm},\omega)_{\hspace*{-0.4mm}{s''}\hspace*{-0.5mm},{s''}}{-}\bar{\tenscomp{N}}^T\hspace*{-1mm}({\bm{q}''})_{\hspace*{-0.4mm}{s''}}\hspace*{-0.5mm}\big),
  \raisetag{8mm}\label{eq:scattering_operator_FT}
\end{split}
\end{equation}

where $\mathcal{V}$ is the unit cell volume, $\bar{\tenscomp{N}}^T\hspace*{-1mm}({\bm{q}})_{\hspace*{-0.2mm}{s}}{=}\big({\exp\hspace*{-1mm}\big[\hspace*{-0.5mm}\frac{\hbar\omega(\hspace*{-0.2mm}\bm{q}\hspace*{-0.2mm})_{\hspace*{-0.2mm}{s}}}{k_B T} \hspace*{-0.5mm}\big]{-}1}\big)^{-1}\hspace*{-1mm}$ is the equilibrium Bose-Einstein distribution, $\Gamma(\bm{q})_{s}{=}(\mathcal{V}N_{\rm c})^{-1}\tenscomp{A}^T(\bm{q},{\bm{q}})_{s,s''}\delta_{s'',s}$ is the phonon linewidth and $\tenscomp{A}^T(\bm{q},{\bm{q}''})_{s,{s''}}$ is the standard collision operator that accounts for anharmonicity and isotopic disorder.
We can now proceed by decoupling the diagonal and off-diagonal equations for $\tenscomp{\tilde{N}}({\bm{K}},{\bm{q}},\omega$). The diagonal expression returns the familiar linearized Boltzmann Transport Equation (LBTE)

\begin{widetext}
\begin{equation}
-i (\omega + 
\vec{\tens{V}}(\bm{q})\cdot\bm{K}) \tenscomp{\tilde{N}}({\bm{K}},{\bm{q}},\omega)_{\hspace*{-0.4mm}{s}\hspace*{-0.5mm},{s}}=\frac{1}{\mathcal{V}N_{\rm c}}\hspace*{-0.5mm}{\sum\limits_{{s''}{\bm{q}''}}}\hspace*{-0.5mm} \tenscomp{{A}}^T(\bm{q},{\bm{q}''})_{\hspace*{-0.4mm}s,{s''}}\hspace*{-0.5mm}\big(\hspace*{-0.5mm}\tenscomp{\tilde{N}}(\bm{K},{\bm{q}''}\hspace*{-1mm},\omega)_{\hspace*{-0.4mm}{s''}\hspace*{-0.5mm},{s''}}{-}\bar{\tenscomp{N}}^T\hspace*{-1mm}({\bm{q}''})_{\hspace*{-0.4mm}{s''}}\hspace*{-0.5mm}\big) + \tenscomp{\tilde{P}}({\bm{K}},{\bm{q}},\omega)_{\hspace*{-0.4mm}{s}\hspace*{-0.5mm},{s}}. \label{eq:LBTE_FT}
\end{equation}
\end{widetext}

The LBTE can be solved through inversion of the collision operator~\cite{chiloyan2017micro}. The off-diagonal expression is 

\begin{equation}
\begin{split}
-i \omega \tenscomp{\tilde{N}}({\bm{K}},{\bm{q}},\omega)_{\hspace*{-0.4mm}{s}\hspace*{-0.5mm},{s'}}
&
+i\Big[\bm{\Omega}(\bm{q}),\tenscomp{\tilde{N}}({\bm{K}},{\bm{q}},\omega)_{\hspace*{-0.4mm}{s}\hspace*{-0.5mm},{s'}}\Big]+\frac{1}{2}\Big\{\vec{\tens{V}}(\bm{q}),\cdot -i\bm{K} \tenscomp{\tilde{N}}({\bm{K}},{\bm{q}},\omega)_{\hspace*{-0.4mm}{s}\hspace*{-0.5mm},{s'}}\Big\}= \\
\frac{\Gamma(\bm{q})_{s}+\Gamma(\bm{q})_{s'}}{2} \tenscomp{\tilde{N}}(\bm{K},\bm{q},\omega)_{s,s'} + \tenscomp{\tilde{P}}({\bm{K}},{\bm{q}},\omega)_{\hspace*{-0.4mm}{s}\hspace*{-0.5mm},{s'}}, \label{eq:offdiagonal_FT}
\end{split}
\end{equation}

which is a simple algebraic expression for $\tenscomp{\tilde{N}}(\bm{K},\bm{q},\omega)_{s,s'}$. Upon inverse Fourier transforming, the solutions to Eqns.~\ref{eq:LBTE_FT} and~\ref{eq:offdiagonal_FT} can be used to obtain expressions for temperature and heat flux

\begin{equation}
    T(\bm{R},t){=}\frac{\hbar}{C \mathcal{V}N_{\rm c}}\sum_{\bm{q},s}\big(\tens{N}(\bm{R},\bm{q},t)\bm{\Omega}(\bm{q})\big)_{s,s}
\label{eq:temperature}
\raisetag{8mm}
\end{equation}

\begin{equation}
    \bm{J}(\bm{R},t){=}
  \frac{1}{2}\frac{1}{\mathcal{V}N_{\rm c}}\sum_{\bm{q},s} {\Big(\big\{ \vec{\tens{V}}(\bm{q}), \tens{N}(\bm{R},\bm{q},t)\big\}\hbar \bm{\Omega}(\bm{q}) \Big)}_{s,s}.
\label{eq:heat_flux}
\raisetag{8mm}
\end{equation}

This approach is general for arbitrary heating profiles in unbounded geometries. The effects of physical boundaries can be captured numerically~\cite{romano2021openbte} or quasi-analytically~\cite{hua2014analytical}. 

\newpage
\bibliographystyle{apsrev}
\bibliography{refs.bib}

\begin{thebibliography}{7}
\expandafter\ifx\csname natexlab\endcsname\relax\def\natexlab#1{#1}\fi
\expandafter\ifx\csname bibnamefont\endcsname\relax
  \def\bibnamefont#1{#1}\fi
\expandafter\ifx\csname bibfnamefont\endcsname\relax
  \def\bibfnamefont#1{#1}\fi
\expandafter\ifx\csname citenamefont\endcsname\relax
  \def\citenamefont#1{#1}\fi
\expandafter\ifx\csname url\endcsname\relax
  \def\url#1{\texttt{#1}}\fi
\expandafter\ifx\csname urlprefix\endcsname\relax\def\urlprefix{URL }\fi
\providecommand{\bibinfo}[2]{#2}
\providecommand{\eprint}[2][]{\url{#2}}

\bibitem[{\citenamefont{Simoncelli et~al.}(2019)\citenamefont{Simoncelli,
  Marzari, and Mauri}}]{simoncelli2019unified}
\bibinfo{author}{\bibfnamefont{M.}~\bibnamefont{Simoncelli}},
  \bibinfo{author}{\bibfnamefont{N.}~\bibnamefont{Marzari}}, \bibnamefont{and}
  \bibinfo{author}{\bibfnamefont{F.}~\bibnamefont{Mauri}},
  \bibinfo{journal}{Nature Physics} \textbf{\bibinfo{volume}{15}},
  \bibinfo{pages}{809} (\bibinfo{year}{2019}).

\bibitem[{\citenamefont{Huberman et~al.}(2017)\citenamefont{Huberman, Chiloyan,
  Duncan, Zeng, Jia, Maznev, Fitzgerald, Nelson, and
  Chen}}]{huberman2017unifying}
\bibinfo{author}{\bibfnamefont{S.}~\bibnamefont{Huberman}},
  \bibinfo{author}{\bibfnamefont{V.}~\bibnamefont{Chiloyan}},
  \bibinfo{author}{\bibfnamefont{R.~A.} \bibnamefont{Duncan}},
  \bibinfo{author}{\bibfnamefont{L.}~\bibnamefont{Zeng}},
  \bibinfo{author}{\bibfnamefont{R.}~\bibnamefont{Jia}},
  \bibinfo{author}{\bibfnamefont{A.~A.} \bibnamefont{Maznev}},
  \bibinfo{author}{\bibfnamefont{E.~A.} \bibnamefont{Fitzgerald}},
  \bibinfo{author}{\bibfnamefont{K.~A.} \bibnamefont{Nelson}},
  \bibnamefont{and} \bibinfo{author}{\bibfnamefont{G.}~\bibnamefont{Chen}},
  \bibinfo{journal}{Physical Review Materials} \textbf{\bibinfo{volume}{1}},
  \bibinfo{pages}{054601} (\bibinfo{year}{2017}).

\bibitem[{\citenamefont{Huberman et~al.}(2019)\citenamefont{Huberman, Duncan,
  Chen, Song, Chiloyan, Ding, Maznev, Chen, and
  Nelson}}]{huberman2019observation}
\bibinfo{author}{\bibfnamefont{S.}~\bibnamefont{Huberman}},
  \bibinfo{author}{\bibfnamefont{R.~A.} \bibnamefont{Duncan}},
  \bibinfo{author}{\bibfnamefont{K.}~\bibnamefont{Chen}},
  \bibinfo{author}{\bibfnamefont{B.}~\bibnamefont{Song}},
  \bibinfo{author}{\bibfnamefont{V.}~\bibnamefont{Chiloyan}},
  \bibinfo{author}{\bibfnamefont{Z.}~\bibnamefont{Ding}},
  \bibinfo{author}{\bibfnamefont{A.~A.} \bibnamefont{Maznev}},
  \bibinfo{author}{\bibfnamefont{G.}~\bibnamefont{Chen}}, \bibnamefont{and}
  \bibinfo{author}{\bibfnamefont{K.~A.} \bibnamefont{Nelson}},
  \bibinfo{journal}{Science} \textbf{\bibinfo{volume}{364}},
  \bibinfo{pages}{375} (\bibinfo{year}{2019}).

\bibitem[{Note1()}]{Note1}
Note1, \bibinfo{note}{a carefully designed experiment that is capable of
  generating populations of distinct phonon modes in a correlated way might be
  one approach.}

\bibitem[{\citenamefont{Chiloyan et~al.}(2017)\citenamefont{Chiloyan, Huberman,
  Ding, Mendoza, Maznev, Nelson, and Chen}}]{chiloyan2017micro}
\bibinfo{author}{\bibfnamefont{V.}~\bibnamefont{Chiloyan}},
  \bibinfo{author}{\bibfnamefont{S.}~\bibnamefont{Huberman}},
  \bibinfo{author}{\bibfnamefont{Z.}~\bibnamefont{Ding}},
  \bibinfo{author}{\bibfnamefont{J.}~\bibnamefont{Mendoza}},
  \bibinfo{author}{\bibfnamefont{A.~A.} \bibnamefont{Maznev}},
  \bibinfo{author}{\bibfnamefont{K.~A.} \bibnamefont{Nelson}},
  \bibnamefont{and} \bibinfo{author}{\bibfnamefont{G.}~\bibnamefont{Chen}},
  \bibinfo{journal}{arXiv preprint arXiv:1711.07151}  (\bibinfo{year}{2017}).

\bibitem[{\citenamefont{Romano}(2021)}]{romano2021openbte}
\bibinfo{author}{\bibfnamefont{G.}~\bibnamefont{Romano}},
  \bibinfo{journal}{arXiv preprint arXiv:2106.02764}  (\bibinfo{year}{2021}).

\bibitem[{\citenamefont{Hua and Minnich}(2014)}]{hua2014analytical}
\bibinfo{author}{\bibfnamefont{C.}~\bibnamefont{Hua}} \bibnamefont{and}
  \bibinfo{author}{\bibfnamefont{A.~J.} \bibnamefont{Minnich}},
  \bibinfo{journal}{Physical Review B} \textbf{\bibinfo{volume}{90}},
  \bibinfo{pages}{214306} (\bibinfo{year}{2014}).

\end{thebibliography}

\end{document}